\documentclass[referee]{cjaa}           % referee version: for submission
\usepackage{graphicx}                   %for PS/EPS graphics inclusion, new
\input{epsf.sty}                        %for PS/EPS graphics inclusion, old
\input{psfig.sty}                       %for PS/EPS graphics inclusion, old

\begin{document}

   \title{Relative spectral lag: an new redshift indicator to measure the cosmos with gamma-ray bursts
%\footnotetext{$*$ Supported by the National Natural Science Foundation of China.}
}
%   \subtitle{I. Place Your Subtitle Here}

   \volnopage{Vol.0 (200x) No.0, 000--000}      %%preserved for Editor. DOn't remove!
   \setcounter{page}{1}          %%starting page, preserved for Editor. DOn't remove!

   \author{Zhi-Bin Zhang \inst{1, 4}
   \mailto{}
%% Please move "\mailto{}" to the corresponding author of the paper.
   \and Jia-Gan Deng \inst{1, 2}
   \and Rui-Jing Lu \inst{1, 2, 4}
   \and Hai-Feng Gao \inst{3}
           }
   \offprints{Z.-B. Zhang}                   %% is disabled in fact

   \institute{Yunnan Observatory, National Astronomical Observatories, Chinese Academy of Sciences,
            P. O. Box 110, Kunming 650011, China\\
             \email{zbzhang@ynao.ac.cn}
%% offprint requests will be sent. Note to pair \mailto{} with \email{}
        \and
        Physics Department, Guangxi University, Nanning, Guangxi 530004, P. R. China\\
%             \email{eloy@iaa.es}
        \and
             Sishui, No. 2 Middle School, Shandong 273206, P. R. China\\
        \and
             The Graduate School of the Chinese Academy of Sciences\\
%             \email{breger@astro.univie.ac.at}
          }

   \date{Received~~2001 month day; accepted~~2001~~month day}

   \abstract{Using 64 ms count data for long gamma-ray bursts ( $T_{90}>$ 2.6 s),
we analyze the quantity, relative spectral lag (RSL), which is
defined as $\tau_{31}/FWHM_{(1)}$, where $\tau_{31}$ is the spectral
lag between energy bands 1 and 3, and $FWHM_{(1)}$ denotes full
width at half maximum of the pulse in energy channel 1. To get
insights into features of the RSLs, we investigate in detail all the
correlations between them and other parameters with a sub-sample
including nine long bursts. The general cross-correlation technique
is adopted to measure the lags between two different energy bands.
We can derive the below conclusions. Firstly, the distribution of
RSLs is normal and concentrates on around the value of 0.1.
Secondly, the RSLs are weakly correlated with $FWHM$, asymmetry,
peak flux ($F_{p}$), peak energy ($E_{p}$) and spectral indexes
($\alpha$ and $\beta$), while they are uncorrelated with
$\tau_{31}$, hardness-ratio ($HR_{31}$) and peak time ($t_m$). The
final but important discovery is that redshift ($z$) and peak
luminosity ($L_{p}$) are strongly correlated with the RSLs which can
be measured easily and directly. We find that the RSL is a good
redshift and peak luminosity estimator.
   }

   \authorrunning{Z. B. Zhang, J. G. Deng, R. J. Lu, et al.}            %author_head in even pages
   \titlerunning{Relative spectral lag: new redshift indicator to measure the cosmos}  % title_head in odd pages

   \maketitle
%% The author head (on even pages) and the title head (on odd pages) will be
%% automatically extracted from \author{} and \title{}. Whenever the title is too long,
%% you will be asked to supply a shorter one by inserting either \authorrunning{} or
%% \titlerunning{} before \maketitle. Anyway, you can specify your own heads in advance.
%%
%%
%% Note: In the following text body of your manuscript, please note several differences from
%%       other major journals:
%% (1) \subsection{Please Capitalize the First Letter of Each Notional Word in Subsection Title}
%% (2) Please Capitalize the First Letter of Each Notional Word in table's caption

%
%________________________________________________ sections below
%
\section{Introduction}           %% first-level sections will be auto-capitalized
\label{sect:intro}
%\hspace{15pt}%                   %% preserved for Editor
The temporal profiles of gamma-ray bursts (GRBs) generally exhibit
very complex and variable characteristics due to overlapping between
adjacent pulses (Norris et al. 1996; Quilligan et al. 2002). So far
many investigations on the analysis of their light-curves especially
the pulses have been made. For example, the properties of pulses
such as widths, amplitudes, area of pulses and time intervals
between them together with number of pulses per burst had been
studied by several authors (e.g. McBreen et al. 1994, 2001, 2003; Li
et al. 1996; Hurley et al. 1998; Nakar \& Piran 2002; Qin et al.
2005).

In addition, some investigations associated with spectra have also
been made (e.g. Kouveliotou et al. 1993; Hurley et al. 1992;
Ghirlanda et al. 2004a). In particular, the spectral lag between
variation signals in different energy bands not only reflects the
features of spectrum evolution but also exhibits the properties of
light-curve. Many researches regarding this variable had been done
from many distinct aspects (see, e.g. Norris et al. 2000, 2001,
2005; Gupta et al. 2002; Kocevski \& Liang 2003; Daigne \&
Mochkovitch 2003; Schaefer 2004; Li et al. 2004; Chen et al. 2005).
It is interestingly found that unlike short GRBs the spectral lags
of most long GRBs are larger than zero and concentrate on the short
end of the lag distribution, near 100 ms ( Band 1997; Norris et al.,
2001).

Concerning the redshift (or luminosity) indictors with GRBs,
previous investigations have offered us some significant paradigms
in the case of light-curves, for instance, the relations of
luminosity-lag (Norris et al. 2000) and luminosity-variability
(Reichart et al. 2001). A particular relation between the lag and
the variability had been strongly confirmed to prove both above
luminosity indicators are reliable (Schaefer et al. 2001). On the
other hand, other indicators based on GRB spectral features are also
constructed subsequently. They originate from either the
$E_{p}-E_{iso}$ relation (Amati et al. 2002; Atteia 2003), the
$E_{p}-L_{p}$ relation (e.g. Yonetoku et al. 2004) or the
$E_{p}-E_{\gamma}$ relation (Ghirlanda et al. 2004b). The spectra
and the light-curves are related to each other, via the spectral
lag.

Norris et al. (2004, 2005) found that wide pulse width is strongly
correlated with spectral lag and these two parameters may be viewed
as mutual surrogates in formulations for estimating GRB luminosity
and total energy. Motivated by the above-mentioned developments, our
first aim is to analyze the RSLs of long bursts in order to see what
the distribution of them should be. Further purpose of this work is
to search for the possible relations between some other parameters
and them, and then interpret them in terms of physics. Data
preparation is performed in $\S$ 2. In $\S$ 3, we are going to
measure some typical physical variables. $\S$ 4 shows all the
results. We are about to apply the RSLs to observed data and try to
reveal their physical explanations in $\S$ 5. We shall end with
discussion and conclusion in $\S$ 6.

%% ChJAA editors DID NOT use \cite{} for citation, \ref and \label for
%% cross-references of Table/Figure in publication version.
%% ChJAA editors prefered you giving a citation as 'Michel et al. 1992', and
%% writting Table~1 or Fig.~1 and so forth. However, that will make authors
%% inconvenient in adjusting/adding/removing text, tables or figures. Anyway,
%% authors can use \cite, \citep and \citet as widely used in other journals.
%% ChJAA editors are moving to use a more flexible LaTeX source.

\section{Data analysis of pulses}
\label{sect:Obs}
%\hspace{15pt}%                   %% preserved for Editor
\subsection{Sample selection}

We use 64 ms count data selected from the current BATSE catalog for
long bursts, called sample 1 including 36 sources. Note that we here
only take into account these bursts with single pulse in the course
of selection. The highly variable temporal structure observed in
most bursts is deemed to be produced by internal shocked outflow,
provided that the source emitting the relativistic flow is variable
enough (e.g. Dermer \& Mitman 1999; Katz 1994; Rees \&
M\'{e}sz\'{a}ros 1994; Piran, Shemi \& Narayan 1993). In this case,
the temporal structure generally reflects the activity of the
``inner engine'' that drives the bursts (Sari \& Piran 1997). As a
result of overlap, it is generally difficult to determine how many
pulses a complex bursts should comprise or to model the shape of
these pulses (Norris et al. 1996; Lee at al. 2000). Fortunately, the
observed peaks have almost one-to-one correlation with the activity
of the emitting source, that is to say, each pulse is permissively
assumed to be associated with a separate emission episode of one
burst (Kobayashi et al. 1997; Kocevski et al. 2003). On the other
hand, spectra parameters for distinct pulses within a burst are
different from each other, which allows us to believe the spectral
lags between these pulses will behave large difference (Hakkila \&
Giblin 2004; Ryde et al. 2005). Therefore we let our sample be
composed of the relatively simple and bright bursts dominated by a
single pulse event here rather than those dim or multi-peak ones for
which we could accurately calculate the spectral lags.

The method of selection is not automated program (e.g. Scargle 1998;
Norris et al. 2001; Quilligan et al. 2002) but directly experienced
eyes, which in a certain degree could reduce any biases either from
denoising techniques or from pulse identification algorithm itself
(Ryde et al. 2003). Lee et al. (2000) has found the numbers of the
pulses within a burst are usually different between energy bands. In
principal, a bright-independent analysis is required as the burst
duration measurement needs (Bonnell et al. 1997), whereas the level
of S/N should be reasonable and reliable. Based on these
considerations, the criterions for our sample selection are now
taken as follows: $T_{90}$ duration $>$ 2.6 s; BATSE peak flux
(50-300 kev) $>$ 1.5 photons $cm^{-2} s^{-1}$; and peak count rate
($>$ 25 kev) $>$ 14000 counts $s^{-1}$. Next, we are going to
process the data via background subtraction and denoising.

\subsection{Background subtraction and denoising}

In general, the first step in data preparation is to select the
appropriate background for substraction. To handle these data as a
whole, an alternative mode of processing involving background
subtraction along with denoising is presented here. For each source,
we take the signal data cover its full range of the pulse as
possible in order to ensure the contributions of all signals to lags
are considered sufficiently. From the point of view of experience,
data beyond this range are regarded as the fit of the background.
However, for convenience, we here prefer disposing of the whole data
involving pre- and post-pulse to processing separately.

Considering the duration and background level of long bursts, we
first smooth them with the DB3 wavelet with the MATLAB software and
then fit them with a pulse function plus a quadratic form, namely
\begin{equation}
F(t)={F_m}(\frac{t}{t_m})^r[\frac{d}{d+r}+\frac{r}{d+r}(\frac{t}{t_m})^{(r+1)}]^{-\frac{r+d}{r+1}}+at^{2}+bt+c
\end{equation}
where the first expression on the right is a quite flexible function
(see eq. (22), Kocevski et al. 2003) applied to describe pulse
shapes and the hinder quadratic term represents a background that
spans the whole data. The parameter $t_m$ is the time of the maximum
flux, $F_m$, of the pulse and the quantities of $r$ and $d$ are next
two indexes describing the rise and decay of pulse profiles
respectively. Once the part of background is subtracted from the
fitted data, the remainders are pure sinal data that isn't
contaminated by background and noise in a certain error level. These
signal data are just what we need to use for the analysis of spectra
and light-curves.

We divide our analyses into two portions in order to achieve
different goals of calculations for sample 1. One portion is to
pretreat the light-curve data in energy channels 1 and 3 (i.e.,
25-55 kev and 110-320 kev). Here, we define these signal data
belonging to the two channels as sample 2. Another is to combine the
data from all four energy channels in order to study the
characteristics of the ``bolometric'' light-curve profile, e.g.
$FWHM$ and asymmetry. To avoid confusion of the definitions, the
summed signal data are called sample 3.
\begin{figure}
\centering
%\vspace{5.5cm}
  \includegraphics[width=2.6in,angle=0]{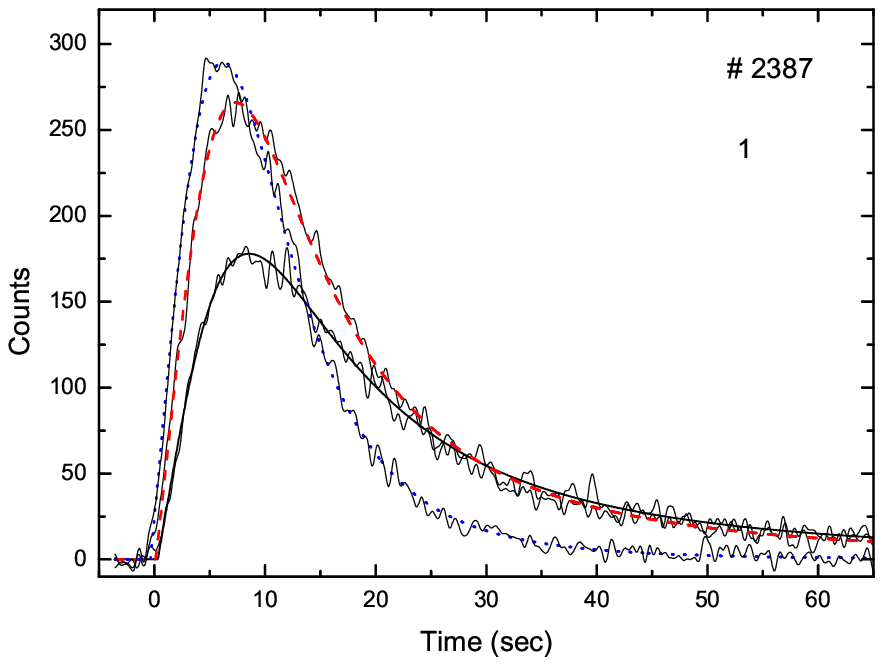}
  \includegraphics[width=2.6in,angle=0]{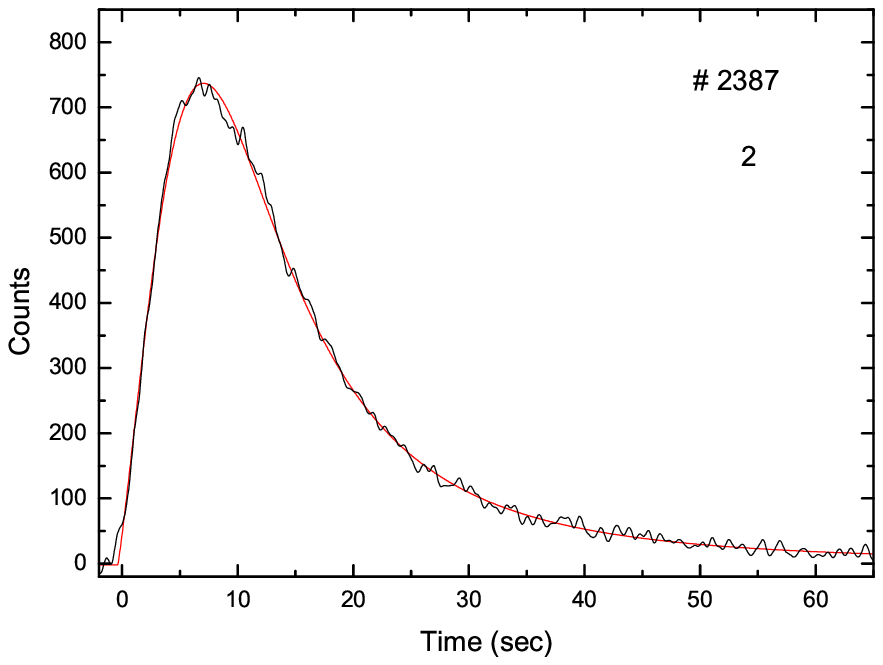}
  \includegraphics[width=2.6in,angle=0]{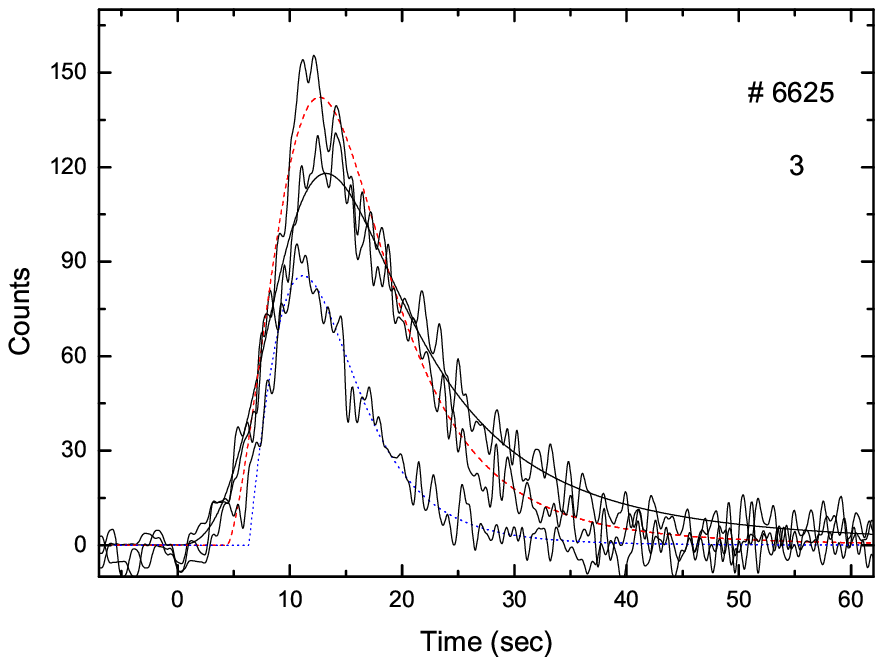}
  \includegraphics[width=2.6in,angle=0]{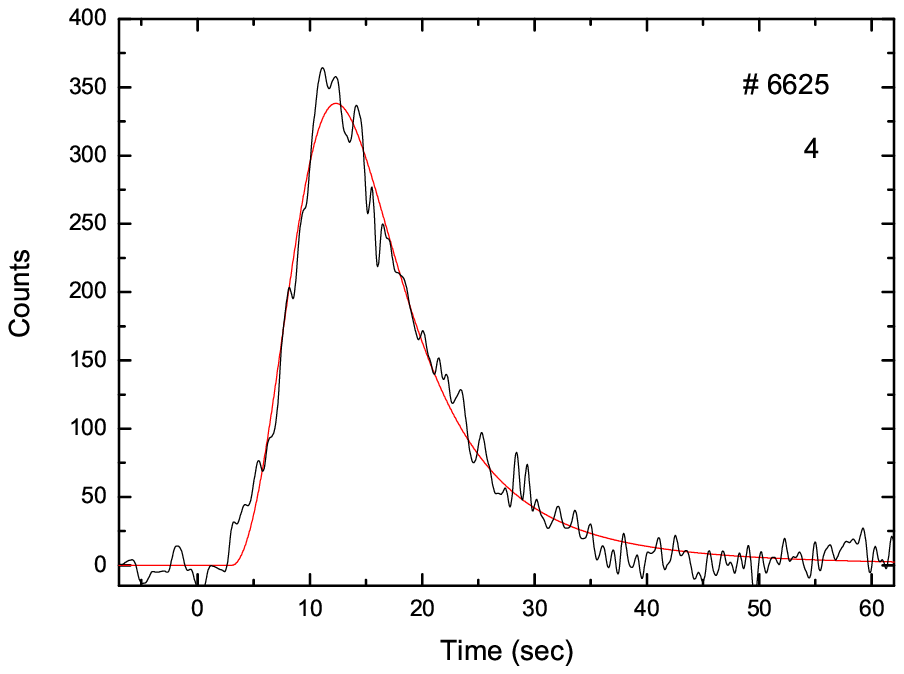}
   \caption{Light-curves of two of studied single-peaked events, GRB
   930612 (\# 2387, panels 1 and 2) with good S/N and GRB 980302 (\# 6625, panels 3 and
   4) with poor S/N. The best fit lines to observed light-curves in energy channels 1 (25-55 Kev), 2 (55-110 Kev) and
   3 (110-320 Kev) are denoted by smooth solid, dashed and dotted lines
   respectively (panels 1 and 3), and the ``bolometric''  light-curves combined from all four energy channels are shown
    with the best fit model (smooth curves) in panels 2
   and 4 respectively.
}
  \label{fig1}
 \end{figure}
Figure 1 displays two examples of the studied pulses (\# 2387 and \#
6625). Panels (1) and (3) show the profiles in energy channels 1, 2
and 3 and their corresponding fit curves identified by solid, dashed
and dotted lines respectively. Panels (2) and (4) depict the the
``bolometric'' light-curve profiles and their fit curves symbolized
by the smooth lines.

\section{Physical Quantities Measure}
\label{sect:data}
%\hspace{15pt}%                   %% preserved for Editor
In this section we focus our attention on measuring some quantities
(RSLs, $FWHM$ and asymmetry) of the single-peaked events from above
signal data for 36 long GRB pulses.

\subsection{Relative spectral lag}

Applying sample 2 we cross-correlate energy bands between energy
channels $j$ and $k$ with the following cross-correlation function
(CCF) (Band 1997)
\begin{equation}
CCF(\tau; \upsilon_{j},
\upsilon_{k})=\frac{<\upsilon_{j}(t)\upsilon_{k}(t+\tau)>}{\sigma_{\upsilon_{j}}\sigma_{\upsilon_{k}}}
\ \ (j\neq k)
\end{equation}
where $\sigma_{\upsilon_{i}}=<\upsilon_{i}^{2}>^{1/2}$, $i$ [=1,2,3
or 4] represents different energy channel; $\tau$ is the so-called
spectral ``lag'' between any two of these channels; $\upsilon_{j}$
and $\upsilon_{k}$ stand for two time series in which they are
respective light curves in two different energy bands. If the
considered channels are appointed to be $j=3$ and $k=1$, the
spectral lag can be thus written as $\tau_{31}$, differing from
those previous definitions of spectral lag (e.g. Norris et al. 2000;
Gupta et al. 2002), we otherwise define a quantity called RSL,
namely
\begin{equation}
\tau_{rel,31}=\tau_{31}/FWHM_{(1)}
\end{equation}
where $\tau_{31}$ represents the lag between energy bands 3 and 1;
and $FWHM_{(1)}$ denotes the full width at half maximum of time
profile in energy channel 1. The $\tau_{31}$ is determined by the
location of $\tau$ where CCF peaks because the CCF curve on this
occasion is smoothing and resembling gaussian shape near its peak.
If the data points close to peak are not dense enough, we shall
interpolate them within the range from one-side to another of the
peak. One could find from this definition that $\tau_{rel,31}$ is
indeed a dimensionless quantity.

\subsection{$FWHM$ \& Asymmetry}

For light-curve of a source, asymmetry and $FWHM$ as the fundamental
shape parameters are especially required to be determined firstly
since they can reflect properties of bursts themselves. The key
factor of this issue may be that $FWHM$ is associated with the
energy of photons detected by observer (or the lorentz factors of
ejecta with $FWHM \propto \Gamma^{-2}$ (see Qin et al. 2004)) and
the asymmetry is largely influenced by co-moving pulse width if only
the burst duration is not large enough, say, $T_{90}<$ 1000s (Zhang
\& Qin 2005). Therefore investigations on asymmetry and $FWHM$
should provide us a probe to detect their intrinsic parameters in
bursts.

We now commence measuring these parameters in a certain significance
level (see $\S$ 3.3). Note that unlike $FWHM_{(1)}$ (in eq [3])
$FWHM$ here represents the full width at half maximum of the
`bolometric' light-curve profile. Asymmetry is defined by the ratio
of the rise fraction ($t_{r}$) of $FWHM$ of pulse to the decay
fraction ($t_{d}$). With a simple algorithm, we quickly measure the
above quantities. Sometimes the sparse data may need to be
interpolated with some points to improve the precision in
calculation. For the purpose of analysis, samples 2 and 3 have been
utilized distinguishingly.

\subsection{Error analysis}

Seen from eqs. (1)-(3), both fitted and derived parameters should be
disturbed by errors propagated between these parameters directly or
indirectly. It's necessary to illustrate the impact of background
portion on observed variables has been eliminated from eq. (1) by
background subtraction. Assuming the error of observed time (t) is
zero at any one of data points, we find that the error of flux
should be caused by parameters $F_m$, $t_m$, $r$ and $d$, i.e.
\begin{equation}
\sigma(F)=\{(\frac{\partial F}{\partial
F_{m}})^{2}\sigma^{2}(F_{m})+(\frac{\partial F}{\partial
t_{m}})^{2}\sigma^{2}(t_{m})+(\frac{\partial F}{\partial
r})^{2}\sigma^{2}(r)+(\frac{\partial F}{\partial
d})^{2}\sigma^{2}(d)\}^{1/2}
\end{equation}
In reverse, $t$ can be expressed as $t\equiv t(x_1, x_2, x_3, x_4,
x_5)=f^{-1}(F, F_m, t_m, r, d)$, so the error of $t$ derived by the
fit to the observed data should be
\begin{equation}
\sigma(t)=\{\sum\limits_{i=1}^5(\frac{\partial t}{\partial
x_i})^{2}\sigma^{2}(x_i)\}^{1/2}=\{\sum\limits_{i=1}^5(\frac{\partial
F}{\partial x_i}/\frac{\partial F}{\partial
t})^{2}\sigma^{2}(x_i)\}^{1/2}
\end{equation}
Suppose two ends of $FWHM$ are $t_1$ and $t_2$ respectively, then
the errors of $t_r$, $t_d$ and $FWHM$ can be written as
\begin{equation}
\sigma(t_r)=[\sigma^{2}(t_1)+\sigma^{2}(t_m)]^{1/2}
\end{equation}
\begin{equation}
\sigma(t_d)=[\sigma^{2}(t_m)+\sigma^{2}(t_2)]^{1/2}
\end{equation}
\begin{equation}
\sigma(FWHM)=[\sigma^{2}(t_1)+\sigma^{2}(t_2)]^{1/2}
\end{equation}
where $t_r=t_m-t_1$, $t_d=t_2-t_m$ and $FWHM=t_2-t_1$. With the
definition of asymmetry ($\equiv t_r/t_d$), its error can be gained
by
\begin{equation}
\sigma(asymmetry)=asymmetry\times[(\frac{\sigma
(t_r)}{t_r})^{2}+(\frac{\sigma (t_d)}{t_d})^{2}]^{1/2}
\end{equation}

The term $\upsilon$ in eq. (2) actually represents the variable $F$
in eq. (1) after background subtraction, which ensures that the
error of $CCF$ is caused by the fluxes fitted to energy channels 1
and 3, that is to say
\begin{equation}
\sigma(CCF)=\{(\frac{\partial CCF}{\partial
\upsilon_1})^{2}\sigma^{2}(\upsilon_1)+(\frac{\partial CCF}{\partial
\upsilon_3})^{2}\sigma^{2}(\upsilon_3)\}^{1/2}
\end{equation}
where $\sigma^{2}(\upsilon_1)$ and $\sigma^{2}(\upsilon_3)$ are
respectively determined by eq. (4) in energy channels 1 and 3. In
addition, the eq. (2) can be in turn replaced by
$\tau_{31}\equiv\tau_{31}(x_1, x_2, x_3)=f^{-1}(CCF,\upsilon_{1},
\upsilon_{3})$. Thus
\begin{equation}
\sigma(\tau_{31})=\{\sum\limits_{i=1}^3(\frac{\partial
\tau_{31}}{\partial x_i})^{2}
\sigma^{2}(x_i)\}^{1/2}=\{\sum\limits_{i=1}^3(\frac{\partial
CCF}{\partial x_i}/\frac{\partial CCF}{\partial \tau_{31}})^{2}
\sigma^{2}(x_i)\}^{1/2}
\end{equation}
It shows that the error of $\tau_{31}$ comes from not only these
fitted parameters but also the $\sigma(CCF)$ itself. The errors of
$\tau_{rel, 31}$ can be deduced from eqs. (2) and (3) by the
following propagation relation as eq. (9)
\begin{equation}
\sigma(\tau_{rel, 31})=\tau_{rel, 31}\times[(\frac{\sigma
(\tau_{31})}{\tau_{31}})^{2}+(\frac{\sigma
(FWHM_{(1)})}{FWHM_{(1)}})^{2}]^{1/2}
\end{equation}
where parameters $FWHM_{(1)}$ and $\sigma(FWHM_{(1)})$ in energy
channel 1 can be measured with the same way as what is used in eq.
(8) as well as eq. (5).

In fact, for a given sample the rise time ($t_r$) and the decay time
($t_d$) have a linear correlation of $t_r \propto \zeta t_d$ where
the value of coefficient $\zeta$ is estimated to be $0.3-0.5$ for
most bursts (Norris et al. 1996). The exact value of $\zeta$ is
uncertain because of the uncertainty in the parameters of the bursts
in the sample. Considering the same reason, the relation between
$\tau_{31}$ and $FWHM_{(1)}$, $\tau_{31}\approx 0.089
FWHM^{-0.42}_{(1)}$ (Norris et al. 2005), is also required to be
calibrated with a larger sample. Otherwise, we can't accurately
determine the errors propagated from these connected parameters. So
we assume they are independent. Under the circumstances, the errors
caused by this un-correlation would become lager than that in the
case of correlation. It's necessary to point out the assumption
doesn't influence the credibility in our error analysis.

\section{Results}

First of all, we display the RSLs distribution of long burst pulses
(\S 4.1). To get the physical implications about it much detailed
investigations have been given to one sub-sample (sample 4), which
includes nine GRBs of long-lag and wide-pulse with simpler physics
owing to more accurate measures (\S 4.2).

\subsection{The RSLs distribution}

Taking $j=3$ and $k=1$ and combining eqs. (2) and (3), we derive the
quantities $\tau_{31}$ and $FWHM_{(1)}$ with samples 2, then
$\tau_{rel, 31}$. Plot of the RSLs distribution is illustrated in
figure 2, from which we can find all the pulses hold positive
$\tau_{rel}$ within the range from 0 to 0.35 and they concentrate on
the approximate value of 0.1.
\begin{figure}
\centering
%\vspace{5.5cm}
 \includegraphics[width=3.6in,angle=0]{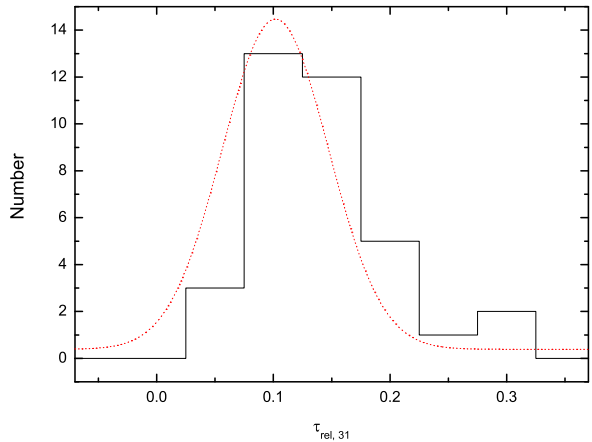}
    \caption{Histogram of the distribution of RSLs for long bursts with a sample of 36 bright pulses.
   The smooth curve is gaussian function fitted to the distribution, where the mean value is $\mu=0.102$ and
   the standard deviation is $\sigma=0.045$.}
  \label{fig2}
 \end{figure}

Moreover, we fit the distribution with gaussian function and get
$\chi^{2}/dof=1.1$ with $R^2=0.97$, which indicates that the
distribution of RSLs is consistent with normal distribution.
However, the distributions of $FWHM$ and spectral lags (or time
intervals) are found to be lognormal instead of normal (see e.g.
Mcbreen et al. 2003). In the following, we pay our particular
attention to the properties of $\tau_{rel, 31}$ that would be
studied in very detail as possible.

\subsection{The dependence of $\tau_{rel, 31}$ on light-curve parameters}

To investigate whether $\tau_{rel, 31}$ is associated with some
parameters of light-curve, we thus list the related parameters in
table 1 and make plots of these relations as shown in figure 3.
\begin{table}[]
  \caption[]{ Parameters gained by fit to sample 4 with the
current model of this paper }
  \label{Tab:publ-works}
  \begin{center}\begin{tabular}{clclclclclclclcl}
  \hline\noalign{\smallskip}
Trigger& $\tau_{rel, 31}^{*} $&$\tau_{31}$ (sec)&$FWHM$ (sec)&Asymmetry&$ HR_{31}$&$F_{m}$&$t_{m}$ (sec)\\
  \hline\noalign{\smallskip}
1406&0.126$\pm$0.017&1.47$\pm$0.15& 9.64$\pm$0.30& 0.43$\pm$0.03& 0.85& 464.09$\pm$2.19&4.47$\pm$0.05 \\
 2387& 0.127$\pm$0.007& 2.43$\pm$0.04& 14.57$\pm$0.30&1.38$\pm$0.07& 0.95&738.89$\pm$2.03&7.42$\pm$0.03\\
 2665& 0.166$\pm$0.014&1.47$\pm$0.10 &4.56$\pm$3.40& 0.44$\pm$0.51&0.45&327.94$\pm$2.49&3.22$\pm$0.26 \\
 3257&0.059$\pm$0.003 &1.61$\pm$0.07 &12.98$\pm$0.74&0.27$\pm$0.02 &1.48&462.97$\pm$2.02 &4.16$\pm$0.06 \\
 6504&0.105$\pm$0.006 &2.02$\pm$0.06&9.78$\pm$1.13 &0.36$\pm$0.07 &1.49&457.78$\pm$3.15&4.39$\pm$0.11 \\
 6625&0.101$\pm$0.04 & 1.55$\pm$0.19&15.51$\pm$1.24& 0.57$\pm$0.08&0.39 &338.18$\pm$2.61&9.24$\pm$0.29\\
 7293&0.092$\pm$0.009 & 1.84$\pm$0.06&12.11$\pm$2.66& 0.31$\pm$0.09& 1.74&626.78$\pm$2.76&5.32$\pm$0.15 \\
 7588&0.149$\pm$0.027 & 1.14$\pm$0.22& 6.56$\pm$0.32&0.56$\pm$0.05 & 0.60& 449.81$\pm$2.63&5.04$\pm$0.06 \\
 7648& 0.192$\pm$0.036& 2.19$\pm$0.30&11.94$\pm$0.60 & 0.55$\pm$0.05& 2.12&305$\pm$4.02&6.89$\pm$0.13\\
  \noalign{\smallskip}\hline
  \end{tabular}
\begin{flushleft}
 Notes: symbol $\ast$ denotes the relative spectral lags calculated with equations (2) and (3), as shown in table 3.
\end{flushleft}
  \end{center}
\end{table}
\begin{figure}
\centering
%\vspace{5.5cm}
 \includegraphics[width=6.2in,angle=0]{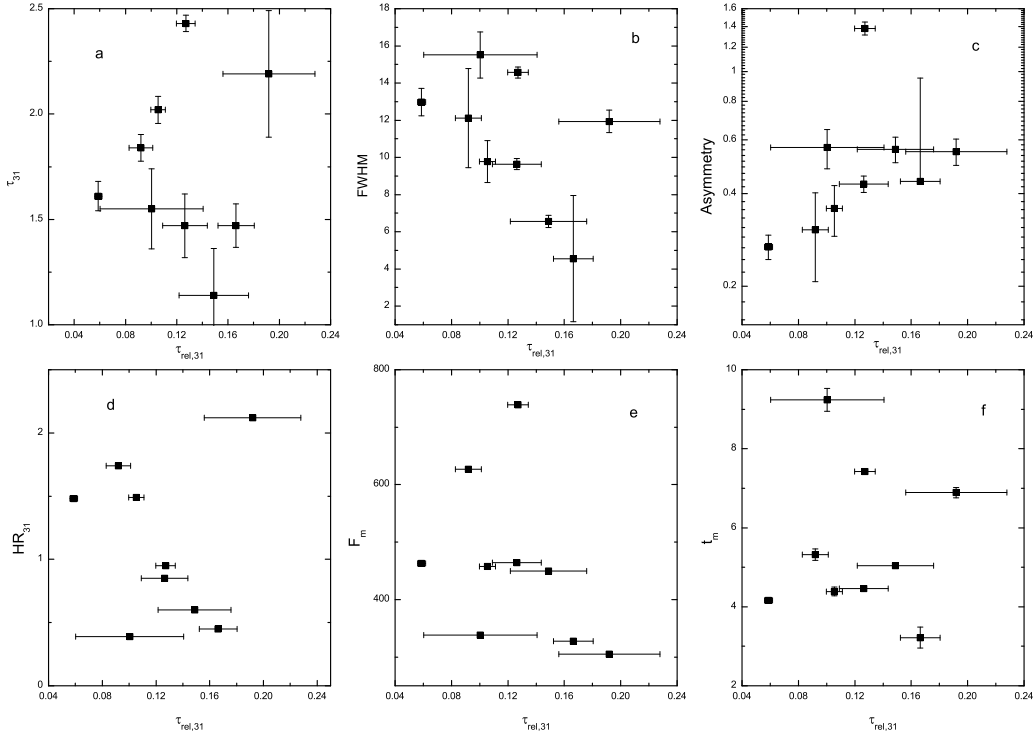}
   \caption{Correlations between $\tau_{rel, 31}$ and other parameters of light-curves such as $\tau_{31}$, $FWHM$,
    asymmetry, $HR_{31}$, $F_m$ and $t_m$. These relations will become more tight provided the source \# 7648 is
    excluded from our sub-sample.}
  \label{fig3}
 \end{figure}
The hardness-ratio ($HR_{31}$) here is defined as the ratio of
photon counts of channel 3 to channel 1 for pure signal data. The
parameters $F_m$ and $t_m$ as well as their errors are taken from
the fit with eq. (1).

For the sake of testing these relations, we take a general linear
correlation analysis (\cite{PR92}) and derive the correlation
coefficients together with probabilities listed in table 2.
\begin{table}[]
  \caption[]{Linear correlation analysis of figures 3 and 4.
Coefficient and probability are abbreviated to coef. and prob.
respectively.}
  \label{Tab:publ-works}
  \begin{center}\begin{tabular}{clclclclclclclcl}
  \hline\noalign{\smallskip}
Panels&&a&b&c&d&e&f\\
  \hline\noalign{\smallskip}
For figure 3&coef.& 0.08&-0.48&0.42&-0.04&-0.39&0.03\\
 &prob.&0.84&0.19&0.25&0.92&0.29&0.95\\
\hline
 For figure 4&coef.& -0.40&0.32&-0.28&-0.89&-0.87&-0.51\\
 &prob.&0.32&0.43&0.49&0.001&0.002&0.16\\
  \noalign{\smallskip}\hline
  \end{tabular}\end{center}
\end{table}
In figure 3, unless the data point (\# 7648) is removed from these
panels (a, d and f), $\tau_{rel, 31}$ is evidently uncorrelated with
$\tau_{31}$, $HR_{31}$ and $t_m$, even though it appears a tendency
of reverse correlation between them. We find from the panels (b, c
and e) that $\tau_{rel, 31}$ shows a positive correlation with
asymmetry, while it is reversely correlated with $FWHM$ and $F_m$.
It demonstrates that the $\tau_{rel, 31}$ is very sensitive to
$FWHM$ and asymmetry, which suggests that it should be another
parameter describing the shape of GRB pulses.

\section{Applications and explanations}

Encouraged by the special attributes of the $\tau_{rel, 31}$, we
shall apply this quantity to explore some other potential
correlations, so that we can have the opportunity to interpret it in
terms of physical significance.

\subsection{Detecting relations between $\tau_{rel, 31}$ and other physical parameters}

First of all, we try to choose the appropriate variables meeting the
aim of studies. Since how to determine the energy spectra as well as
distances (or energies) is usually considered to be the key issue of
understanding the burst phenomenon, we therefore select the relevant
parameters such as $\alpha$, $\beta$, $E_{p}$, $z$, $L_{p}$ and
$F_{p}$ to correlate with $\tau_{rel, 31}$ respectively (see table
3).
\begin{table}[]
  \caption[]{ Parameters for observed and modeled data in
sample 4 }
  \label{Tab:publ-works}
  \begin{center}\begin{tabular}{clclclclclclclcl}
  \hline\noalign{\smallskip}
Trigger& $\tau_{rel,31}^{\ast}$&$\alpha^{a}$&$\beta^{a}$&$E_{p}^{a}$&$z^{b}$&$L_{p}^{b}$&$F_{p}^{a}$\\
\hline\noalign{\smallskip}
1&2&3&4&5&6&7&8\\
  \hline\noalign{\smallskip}
1406&0.126$\pm$0.017& -1.20$\pm$0.15& -2.89$\pm$0.21& 79.43$\pm$21.39& 1.91$\pm$0.4& 22.8$\pm$10 &2.13\\
 2387&0.127$\pm$0.007& -0.03$\pm$0.09& -2.46$\pm$0.03&106.79$\pm$8.00 &3.76$\pm$0.4 & 176$\pm$50& 3.58\\
 2665&0.166$\pm$0.014 &------&------ &------ &1.19$\pm$0.11 & 11$\pm$2& 1.91\\
 3257& 0.059$\pm$0.003&-0.24$\pm$0.09 &-2.79$\pm$0.12 &169.2$\pm$16.32&11.97$\pm$1.6 &1750$\pm$600 & 2.62\\
 6504& 0.105$\pm$0.006&0.59$\pm$0.26 &-2.78$\pm$0.17 &120.2$\pm$20.34 &1.67$\pm$0.07 &40$\pm$4& 2.25\\
 6625&0.101$\pm$0.04 & -1.18$\pm$0.12& -4.54$\pm$1.76&53.8$\pm$9.31&2.18$\pm$0.4 &15.8$\pm$6 &1.68 \\
 7293& 0.092$\pm$0.009&-0.15$\pm$0.12& -2.81$\pm$0.11& 151.1$\pm$17.41& 8.48$\pm$3&733$\pm$500 &2.73 \\
 7588&0.149$\pm$0.027&-1.73$\pm$0.38&-2.80$\pm$0.08 & 12.07$\pm$17.9& 1.44$\pm$0.08&7.79$\pm$2.2 & 2.06\\
 7648& 0.192$\pm$0.036& -0.78$\pm$0.26&-2.43$\pm$0.21& 146.93$\pm$58.0& 0.43 $^{c}$& 0.54$\pm$0.1 $^{d}$
& 1.53\\
  \noalign{\smallskip}\hline
  \end{tabular}
\begin{flushleft}
Notes.-- Redshift (col. 6) and peak luminosity (col. 7) estimated by
the $E_{p}-L_{p}$ relation have been borrowed from Yonetoku et al.
(2004) due to lack of the information about these sources except for
trigger 7648 whose $z$ and $L_P$ ($10^{51}$ergs s$^{-1}$) is offered
by Galama et al. (1999) and Guidorzi et al. (2005)
respectively. Note the unit of peak flux ($F_{p}$) is photons $cm^{-2} s^{-1}$.\\
References.-- a. Norris et al. 2005; b. Yonetoku et al. 2004; c.
Galama et al. 1999; d. Guidorzi et al. 2005.
\end{flushleft}
  \end{center}
\end{table}

It needs to clarify that the redshifts of all sources except \# 7648
are borrowed from their estimation by an empirical $E_{p}-L_{p}$
relation instead of direct observations for the absence of their
information of the afterglow. The reason for this selection is
$E_p-L_p$ relation looks considerably tighter and more reliable than
the relations offered by the previous works (\cite{yo04}), whereas
the spectral parameters ($\alpha$, $\beta$ and $E_p$) for \# 2665
are unavailable at present.

All the correlations of $\tau_{rel, 31}$ with these parameters are
shown in figure 4.
\begin{figure}
\centering
%\vspace{5.5cm}
 \includegraphics[width=6.2in,angle=0]{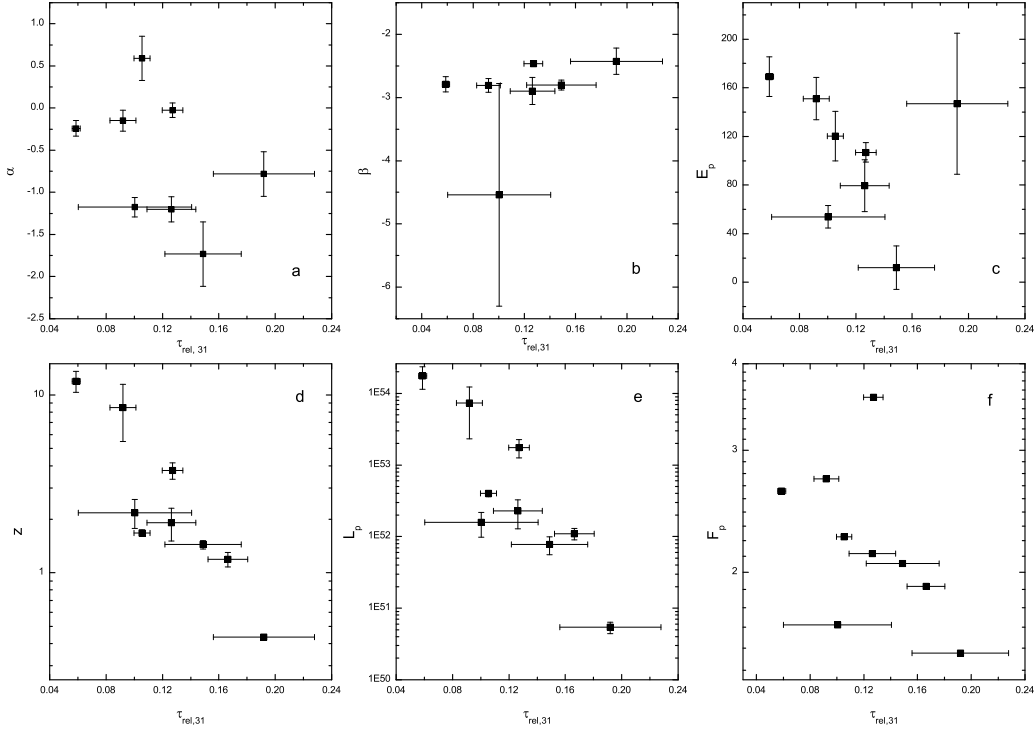}
    \caption{Correlations between $\tau_{rel, 31}$ and other parameters such as $\alpha$, $\beta$, $E_{p}$, $z$,
$L_{p}$ and $F_{p}$. The relations also have the tendency of
becoming more tight than before once the source \# 7648 is
    excluded from our sub-sample.}
  \label{fig4}
 \end{figure}
In the same way, we apply the linear correlation analysis to these
relations and list the corresponding results in table 2. Figure 4
illustrates all parameters but $\beta$ are reversely correlated with
$\tau_{rel, 31}$. The high energy spectral index, $\beta$, behaves
an otherwise trend of positive correlation with $\tau_{rel, 31}$. It
has been proposed that $F_{p}$ is often used as an effective
indictor of distance to the GRB sources (Lee et al., 2000). The
tight relation between $\tau_{rel, 31}$ and $F_{p}$ in figure 4(f)
shows that $\tau_{rel, 31}$ is expected to be a distance indictor.

It is surprisedly found from panels (d) and (e) that $\tau_{rel,
31}$ is strongly correlated with $z$ and $L_p$ respectively.
Accordingly, our attempt in next section is to confirm if
$\tau_{rel, 31}$ is suitable for an indictor of distance or peak
luminosity.

\subsection{Redshift and luminosity indicator}

As Atteia (2005) points out, whether redshift indictors are good or
not is determined by the degree of correlation between redshift and
them which are generally combinations of GRB parameters with a small
intrinsic scatter. To testify the validity of $\tau_{rel, 31}$ as
the redshift indicator, we contrast the observed data with the
theoretical model in figure 5. Note that the data of panels (a) and
(b) are merely a replica of figure 4(d) and (e) correspondingly.
\begin{figure}
\centering
%\vspace{5.5cm}
 \includegraphics[width=3.6in,angle=0]{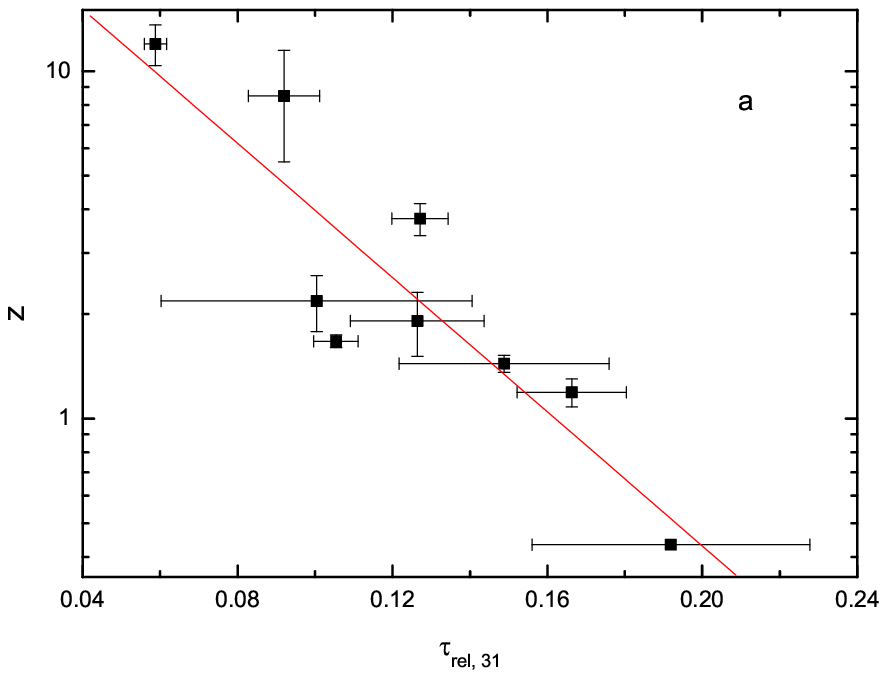}
 \includegraphics[width=3.6in,angle=0]{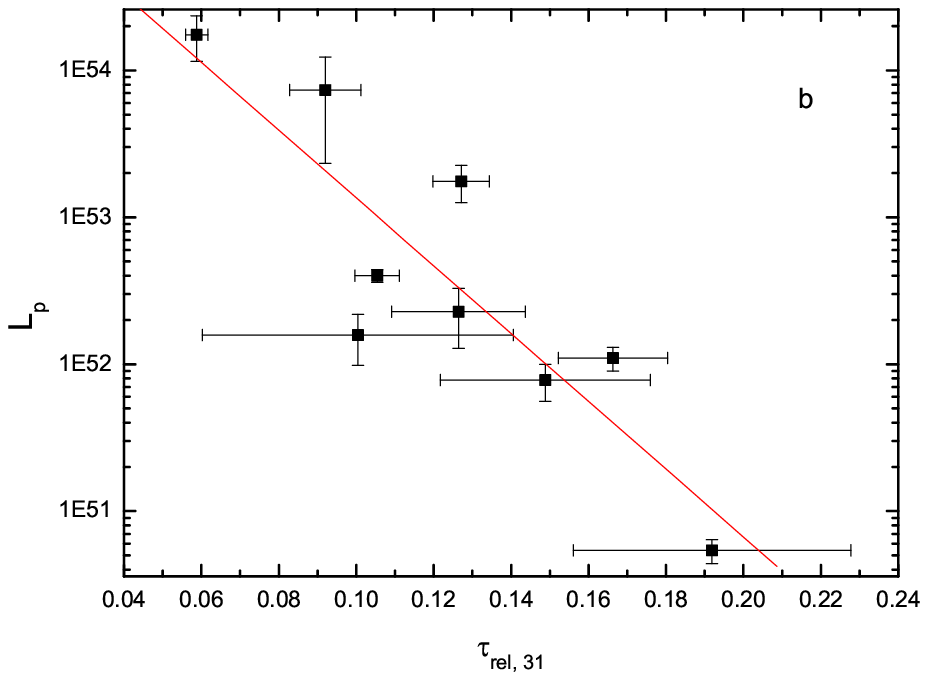}
    \caption{Calibration curves for the relative spectral lags, $\tau_{rel, 31}$. Plots of $\tau_{rel, 31}$ vs. redshift
    and peak luminosity can be used to calibrate redshift and/or luminosity indicators. The plots here can be
    fitted to yield eqs. (13) and (14) (marked with the straight red lines in panels a and b).}
  \label{fig5}
 \end{figure}
From figure 5(a) and (b), the best fits to a linear model are
written as
\begin{equation}
\log z \approx1.56-9.66 \tau_{rel, 31}
\end{equation}
\begin{equation}
\log L_p \approx55.44-23.07 \tau_{rel, 31}
\end{equation}
with spearman rank-order correlation coefficients of -0.88 ($p\sim
1.5\times10^{-3}$) for the former and -0.83 ($p\sim 5\times10^{-3}$)
for the latter. This indicates the relatively accurate connection
between the RSLs and the redshift (or luminosity) does exist.
Provided the $\tau_{rel, 31}$ is measured, using eqs. (13) and (14)
one could precisely estimate redshifts and peak luminosities of
those sources without the information of observed spectral lines.
From this viewpoint, the quantity $\tau_{rel, 31}$ can be regarded
as an ideal indicator of redshift and/or luminosity.

Meanwhile, the RSLs might be utilized to constrain the cosmological
parameters (say, $\Omega_{m}$, $\Omega_{\lambda}$ and $H_0$) in case
redshift and luminosity are determined by eqs. (13) and (14)
simultaneously. Certainly, the realization of this purpose requires
us to eliminate selection effects as possible as we can in advance,
not only on observations but also on calculations.

\section{Conclusions and discussions}

According to above investigations, we could derive the following
conclusions: First, the distributions of RSLs are normal and
concentrate on the value of 0.1 or so. Second, the RSLs are weakly
correlated with $FWHM$, asymmetry, $\alpha$, $\beta$, $E_{p}$,
$F_{p}$ and $F_m$. While they are uncorrelated with $lags_{31}$,
$HR_{31}$ and $t_{m}$. Finally, we find the $\tau_{rel, 31}$ is a
useful redshift and peak luminosity estimator.

In view of internal shock model, GRBs generally consist of many
pulses produced by multiple relativistic shells (or winds) followed
by internal shocks due to collision of them as a central engine
pumps energy into medium (e.g. Fenimore et al. 1993; Rees \&
M\'{e}sz\'{a}ros \ 1994). Even so, our action on sample selection
doesn't conflict with the idea of overlapping. For simplicity, we
only choose single peaked bursts to construct our sample to make
this issue easier to study. Furthermore, the manner of selection can
also avoid the contamination of adjacent pulses by overlap, for
which could inevitably produce additional errors, owing to selection
effects. Under the assumption that these single wide pulses could be
produced by the same mechanism (see Piran 1999), our motivation of
this selection is advisable. In fact, McMahon et al. (2004) has
pointed out the favorite mechanism for producing gamma-ray emission
for such single pulse events could be not internal shocks but
external shocks (see also M\'{e}sz\'{a}ros \& Rees 1993; Sari \&
Piran 1996; Dermer et al. 1999, 2004).

It is usually expected that the gamma-rays come from a
relativistically expanding fireball surface with lorentz factor
$\Gamma>100$ (e.g. Lithwick et al. 2001). The bulk Lorentz factor
increases linearly with radius as long as the fireball is not baryon
loaded and not complicated by non-spherical expansion (Eichler et
al. 2000) until $\Gamma>1000$ (or $R\geq10^{16} cm$) (Woods et al.,
1995). With Power density spectrum method, Spada et al. (2000) found
the curvature time together with the dynamic time will dominate over
the radiative cooling time at a distance $R<5\times10^{14}cm$. Given
this case, based on Doppler effects the $FWHM$ has been found to
follow $FWHM\propto\Gamma^{-2}$ (Qin et al. 2004). Assuming $\tau
\propto \Gamma^{-\omega}$, from figure 3(b) we can deduce that the
upper limit of $\omega$ is about 2 which is compatible with $\tau
\propto\Gamma^{-1}$ gained by Shen et al. (2005).

In this work we have shown that relative spectral lag can be used as
an estimator of redshift and peak luminosity of long GRBs. We note
that our conclusion is based on an analysis using nine sources. More
accurate and robust results for the analysis would require a larger
sample which includes sources with redshifts. Until now the expected
indictor of short bursts hasn't been constructed yet, owing to lack
of sufficient sources with the redshift, although redshifts of few
sources have been measured by Swift and HETE II and reported
recently (see, e.g. Bloom et al. 2005; Hjorth et al. 2005; Berger et
al. 2005). The recent observations suggest that short bursts reside
at cosmological distances, however, previous investigations showed
the spatial population of short GRBs seems to accord with lower
redshift sources (e.g. che et al. 1997; Magliocchetti et al. 2003;
Tanvir et al. 2005). Therefore, the research on redshift indictor of
short bursts is excluded from this work and it still remains at the
prenatal stage.

\begin{acknowledgements}

We would like to thank the anonymous referee for many helpful
comments. It's my honor to thank D. Kocevski for useful suggestions
in the preparation of this manuscript. This work was supported by
the Special Funds for Major State Basic Research Projects (973) and
National Natural Science Foundation of China (No. 10273019 and No.
10463001).

\end{acknowledgements}

\appendix                  %%appendicial material is supported

\label{lastpage}

\end{document}